\documentclass[twocolumn,prl,floatfix]{revtex4}

\usepackage{subfig}
\usepackage{graphicx}
\usepackage{type1cm}
\usepackage{eso-pic}
\usepackage{color}
\usepackage{amssymb}

\begin{document}

\title{The performance of Minima Hopping and Evolutionary Algorithms for cluster structure prediction}
\author{Sandro E. Sch\"{o}nborn, Stefan Goedecker, Shantanu Roy}
\affiliation{Departement Physik, Universit\"{a}t Basel, Klingelbergstr. 82, 4056 Basel, Switzerland}
\author{Artem R. Oganov}
\affiliation{Laboratory Of Crystallography, Department Of Materials, ETH Z\"{u}rich, 8093 Z\"{u}rich, Switzerland\\
             Geology Department, Moscow State University, 119899 Moscow, Russia}

\keywords{global optimization, evolutionary algorithm, minima hopping,
atomic cluster, Lennard-Jones cluster, silicon cluster, gold cluster}

\begin{abstract}
We compare Evolutionary Algorithms with Minima Hopping for global
optimization in the field of cluster structure prediction. We introduce
a new {\em average offspring} recombination operator and compare it
with previously used operators. Minima Hopping is improved with a {\em
softening} method and a stronger feedback mechanism. Test systems are
atomic clusters with Lennard-Jones interaction as well as silicon and
gold clusters described by force fields.  The improved Minima Hopping
is found to be well-suited to all these homoatomic problems.  The
evolutionary algorithm is more efficient for systems with compact and
symmetric ground states, including LJ$_{150}$, but it
fails for systems with very complex energy landscapes and asymmetric
ground states, such as LJ$_{75}$ and silicon clusters with more than
$30$ atoms.  Both successes and failures of the evolutionary
algorithm suggest ways for its improvement.
\end{abstract}

\maketitle

\section{Introduction}

To find the structural ground state of a cluster is a non-trivial global
optimization task. One has to find the global minimum of the potential
energy surface which is a function of all the atomic coordinates.  Even
for a relatively small cluster of 30 atoms the configuration space has
already 90 dimensions. Because knowing the structure is a prerequisite
for the study of all other physical and chemical properties the problem
is of great importance and many algorithms have been developed to
solve this global optimization problem.  We compare {\em Evolutionary
Algorithms} which have successfully been used in many
diverse fields with the {\em Minima Hopping}~\cite{Goedecker2004} method.

For the prediction of the ground state structure of crystals,
the {\em USPEX (Universal Structure Predictor:  Evolutionary
Xtallography)}~\cite{Oganov2006,Glass2006} method has turned out
to be extremely powerful and has already allowed material
scientists to find interesting and unexpected new crystal
structures~\cite{Gao2008,Oganov2006a,Oganov2008,Oganov2007,Ma2007}.
Recently Evolutionary Algorithms have also been
successfully used to predict surface phenomena such as steps on
silicon crystals~\cite{Chuang2005,Briggs2007}.  A widespread
application of global optimization methods is the prediction of
the structure of various clusters, In this field the majority of
the work has been done with genetic or evolutionary methods as
well~\cite{Deaven1995,Deaven1996,CCD,Johnston2003}. We note that
different evolutionary algorithms developed for various types of
structure prediction problems (molecules, clusters, crystals) have
significant differences. Even for the same type of problem (e.g. crystal
structure prediction) the previously proposed algorithms are very
different in their construction and performance.

The minima hopping method has been successfully applied to benchmark
systems~\cite{Goedecker2004} as well as to silicon clusters and AFM
tips~\cite{Ghasemi2008}.

The presented evolutionary algorithm is similar to and inspired by
the structure prediction algorithm USPEX. But since we work with
non-periodic systems without a crystal lattice, some modifications of the
original method were required.  The version of minima hopping we are
using is based on an improvement of the two key features of the original
minima hopping method~\cite{Goedecker2004}.  The feedback mechanism
is enhanced and the Bell-Evans-Polanyi principle~\cite{Roy2008} is
exploited in a more efficient way by moving preferentially along soft
directions in the molecular dynamics (MD) part of the minima hopping algorithm.

Our comparison of Minima Hopping and Evolutionary Algorithms is based
on Lennard-Jones systems, especially the cluster with $55$ atoms, which
is an example of an easy one-funnel structure, and the $38$-atom system,
which is known to have a complicated double-funnel structure. We also
apply the algorithms to more realistic systems, namely silicon clusters
described by a force field and gold clusters described by an embedded
atom potential.

This paper is structured as follows: We first introduce the evolutionary
method used, after a quick introduction to Minima Hopping and its
modifications we present the results section containing a comparison
between Minima Hopping and the Evolutionary Algorithm and finally we
also test different aspects of the EA and MH.

\section{The Evolutionary Algorithm}

Evolutionary Algorithms (EA) implement a very simple model of biological
evolution. They work on a set of samples --- a \emph{population} ---
which is gradually improved by selection and reproduction of fit members
of the population --- \emph{individuals}. Each individual is a solution
candidate.  A single iteration step leads from a population to the
next and is called a \emph{generation}. The algorithm optimizes the
\emph{fitness} function --- in our context the negative energy of the
configuration.  The operators applied to the population to obtain the
next generation are the heart of the algorithm as they determine its
quality and properties.

In contrast to the original simple genetic algorithm, modern
applications in the field of chemical structure prediction
all use real value encoding instead of binary strings and
phenotypical operators acting directly in real space instead of gene
modification~\cite{Pullan1997,Johnston2003}. Also state of the art is
the application of local optimization to each individual thus reducing
the search space to basin bottoms. Local optimization is done using standard
techniques such as steepest descent and conjugate gradient methods.

\begin{figure}[ht]
   \begin{center}
        \includegraphics[width=8cm]{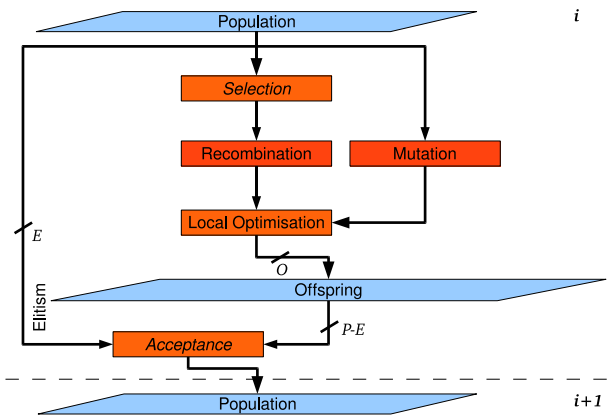}
    \end{center}
    \caption{Evolutionary Algorithm: A population
    $i$ is evolved into the next generation $i+1$ using mutation and
    recombination operators. $E$: elitism (number of individuals
    kept from the old generation). $O$: total number of offspring
    produced. $P$: population size.}
    \label{fig:flowchart}
\end{figure}

FIG.~\ref{fig:flowchart} presents an overview of the Evolutionary
Algorithm used.  It starts with a randomly initialized population, this
is our generation $0$. In each generation the algorithm goes through
three steps: \emph{Selection}, \emph{Application of Operators} and
\emph{Acceptance}. The steps {\em Selection + Operator} produce new
offspring and mutations and put them into an intermediate population.
In {\em Acceptance} the next generation is selected out of all available
offspring together with individuals from the old population.

The algorithm possesses many tunable parameters. Most important
values are {\tt populationsize}, the number of {\tt offspring}
produced, the number of individuals kept from the last generation
({\tt elitism}) and the {\tt mutationrate}, for a complete list see
TABLE~\ref{tbl:parameterlist}.

\begin{table*}
    \centering
    \begin{tabular}{l|l|r}
        \hline
        \hline
        {\bf Function}            & {\bf Name}                                                          & {\bf Standard Value} \\
        \hline
        Population size                                               & {\tt populationsize}& $30$ \\
        Number of offspring produced                                  & {\tt offspring}     & {\tt populationsize} \\
        Number of individuals taken from former population            & {\tt elitism}       & $1/3$ {\tt populationsize} \\
        Last rank with selection probability $>0$                     & {\tt cutoff}        & $2/3$ {\tt populationsize} \\
        Relative rate of offspring produced with {\em average} method & {\tt avgoffspring}  & $0.50$ \\
        Only one individual allowed within this energy interval       & {\tt energyslot}    & $10^{-4}$ \\
        Total rate of mutation                                        & {\tt mutationrate}  & $0.05$ \\
        Random walk mutations (relative to total mutations)           & {\tt mutrwalk}      & $0.60$ \\
        Strain mutations (relative to total mutations)                & {\tt mutstrain}     & $0.30$ \\
        Probability of random rotation before recombination           & {\tt raterndrot}    & $0.90$ \\
        Convergence criterion for force norm in local optimizer       & {\tt fnrmtol}       & $10^{-4}$ \\
        \hline
        \hline
    \end{tabular}
    \caption{Parameters of the Evolutionary Algorithm}
    \label{tbl:parameterlist}
\end{table*}

As it is often the case in evolutionary algorithms in this field an
energy slot restriction allows only one candidate per energy interval
{\tt energyslot} in the population. This method of preventing multiple
copies of the same configuration in the population may be dangerous for
it might reject an important candidate having almost the same energy as
an individual already known. Using force
fields allows to calculate energy and forces with very high precision
since the numerical noise is extremely low. It
is thus easily possible to identify structures by their energy.

%<+shift to end?+>
Recently, a different structure of EA, optimized for parallel machines,
has been presented~\cite{Bandow2006}.  Instead of a stepwise evolution
this approach handles a big pool of individuals which is subject to
continuous application of operators. This is closer to a biological
population without sharply defined generation gaps and it solves the
load balancing problem in parallel implementations of EA.

\subsection{Operators}

Operators are used to evolve a population to a next generation.  We use
two different kinds of operators, {\em heredity} operators which take
two individuals as input and produce a {\em child} sharing properties of
both {\em parents}. The second kind is an operator applied to a single
individual altering its configuration ({\em mutation}).

The selection step determines to which individuals the operators are
applied, it is dependent on the operator. For a heredity operator
there are two parents selected whereas for mutation operators only
one individual is chosen. Selection is done using a linear ranking
scheme.
Individuals are sorted with respect to their fitness values and
then assigned a probability depending linearly on the rank $i$. 
The probability of selecting the individual with rank $i$ is in this case 
\begin{equation}
P[i] = P_1 - (i-1) \frac{P_1}{c}
\label{eqn:rankprob}
\end{equation}
where $i$ is the rank, starting at $1$, $c$ the parameter {\tt
cutoff} and $P_1$ the first selection probability determined by
normalization constraint. The cut-off value is the last rank with a
selection probability above zero, all following ranks are assigned
zero selection probability. The same method is also used in USPEX
method~\cite{Oganov2006,Glass2006}, it has turned out to be more
efficient than Boltzmann selection where the selection probability
follows a Boltzmann distribution depending on relative fitness
values.  The selection of the same individual serving as both parents
is prevented. Mutation operators are applied randomly to the whole
population.

The first heredity operator used is the cut and splice
(\emph{cutting-plane}) method introduced by Deaven and
Ho~\cite{Deaven1995,Deaven1996}. Both clusters are centered at their
center of mass (COM). A randomly oriented plane cuts the two clusters
apart. The new {\em offspring} cluster consists of one half of the first
and the other half of the second cluster. Though able to obtain two
offspring by this method we only produce one. The plane usually contains
COM and its cut preserves the total number of atoms in the child cluster.

A new way of producing offspring is implemented in the \emph{average
offspring} method. Both clusters are centered at COM and for each atom
of the first cluster the closest lying atom of the other cluster is
identified. The atom of the child is now placed randomly on the connecting line of the two parent
atoms. The randomness of this operator is necessary to prevent producing
a lot of identical offspring.

Before application of either heredity operator a random rotation can be
performed on one of the parents.  The frequency at which this rotation
is used can be adjusted via {\tt raterndrot}.

Mutations are introduced to keep the diversity of the population
high and prevent premature convergence. Three different methods
are used. The easiest of those is the \emph{random walk mutation}
where atoms are randomly displaced. The displacement is approximately
normal distributed with a mean displacement in the order
of the two-body potential equilibrium distance (bond length). A
\emph{strain mutation} applies a geometrical deformation to the
whole cluster, inspired by mechanical stress. The deformations
include (anisotropic) compression and shear. Such strain transformations 
gave increased efficiency in the USPEX method. If the compression is high this method
relates to the \emph{Big-Bang-Algorithm} where configurations are
relaxed from very high compressions~\cite{Leary1997}. We only apply
a moderate compression.  In a third type of mutation the cluster is
cut into two pieces similar to the plane-cutting method. One of the
pieces is rotated around an axis perpendicular to the cutting plane
by a random angle. This method was introduced as \emph{twinning
mutation}~\cite{Wolf1998}.

Since our operators are able to produce configurations with atoms
lying very close to each other we use a pre-relaxation method which is
essentially a steepest descent with a very small step size.
After a few steps the real self-adjusting steepest descent is started until a
nearly quadratic region around the local minimum is reached. The final
minimization is then carried out by a conjugate gradient method.

If an intermediate set of offspring has been created the {\em Accepting}
step is triggered.  In this step it is decided whether an offspring
is accepted into the new population or is discarded.  The algorithm
first accepts the best individuals from the former population ({\em
elitism}). The number of individuals chosen that way is given by {\tt
elitism}.  In a second step the individual with the worst fitness
value is replaced by the best offspring if energy slot constraints are
fulfilled.  This is repeated until all offspring are processed or the
population is complete.

The algorithm is left running until a given limit of generations has
been reached or the (known~\cite{CCD}) global minimum has been found.

\section{Minima Hopping}

Minima Hopping is a recently developed global optimization
algorithm~\cite{Goedecker2004} which makes use of a
Bell-Evans-Polanyi (BEP) principle for Molecular Dynamics (MD)
trajectories~\cite{Roy2008}. The BEP principle states that low energy
MD trajectories are more likely to enter the basin of a lower lying
adjacent minimum than high energy trajectories. The algorithm also
incorporates a history to repel it from previously visited regions.
Using local optimization and molecular dynamics simulation it jumps
between basins of attraction.  The kinetic energy is kept as low as
possible to escape the local minimum but is increased if this minimum
has already been visited before.

Minima Hopping works with two self-adapting parameters, the kinetic energy
of a MD escape step {\tt ekin} and an acceptance threshold
{\tt ediff} for new minima to introduce a further downward
preference. Starting from a local minimum a MD escape trial is
started with kinetic energy {\tt ekin}. After a few steps it
is stopped and the configuration locally optimized again.
If escaped the new minimum is only
accepted when the new energy lies at maximum {\tt ediff} higher
than that of the previous minimum.

Minima Hopping is adjusted by tuning five feedback parameters: 
$\alpha_1$ decreases {\tt ediff} when a new minimum is accepted
whereas $\alpha_2$ increases the threshold on rejection, $\beta_1$ 
increases the kinetic energy when a
MD escape trial fails, $\beta_2$ increases {\tt ekin} when the new minimum
is already known and $\beta_3$ decreases the kinetic energy if the
minimum is unknown. Each visited minimum is added to a history list and
marked as known. The algorithm currently uses the energy value to identify different minima.
For more details we refer to the original paper~\cite{Goedecker2004}.

Minima Hopping was used with the standard parameter set presented in the original
paper: $\alpha_1=1/1.05, \alpha_2=1.05$ and $\beta_1, \beta_2=1.05, \beta_3=1/1.05$.

The algorithm is efficient since it inhibits revisiting the
same configuration many times which is likely to be the case in
thermodynamically inspired methods such as {\em Simulated Annealing}.

%The starting configuration of this algorithm is chosen to be a sphere
%cut out of a simple cubic lattice. As no ground state in this work is
%based on a simple cubic lattice it is acceptable to start from this
%configuration. Internal comparisons with other starting configurations have
%shown that this starting point is comparable in performance with other
%starting configurations, such as fcc lattice or a simple cubic box, even
%with random initialisation.

We present two modifications of the original algorithm:
The initial velocity vector of a MD escape trial is moved
towards a direction with low curvature ({\em softening}) and a
stronger feedback mechanism is used.

%<+SOFTENING+>
\paragraph{Softening.}
MD escape trials in the MH algorithm need an initial velocity distribution which is then
rescaled to fit the desired kinetic energy. The velocities are randomly
directed for each atom with Gaussian distributed magnitudes.
Regardless of the actual distribution chosen it has proved very useful
to use {\em softening}, to choose velocities along low-curvature directions. 
In this way one can typically find MD trajectories with a relatively small 
energy that cross rapidly into another basin of attraction.
In the original MH method low kinetic energy trajectories could only be obtained by 
using large values for {\tt mdmin} which results in long trajectories.
A direction of low curvature is found using a modified iterative dimer method
which only uses gradients, no second derivatives need to be calculated~\cite{Henkelman1999}.\\
Starting at a local minimum $\mathbf x$ with an escape direction $\mathbf{\hat{N}}$
the method calculates a second point $\mathbf{y} = \mathbf{x} + d \mathbf{\hat{N}}$
at a distance $d$ along the escape direction.
The forces are evaluated at $\mathbf y$ and the point is moved along a
force component $\mathbf{F}^\perp$ perpendicular to $\mathbf{\hat{N}}$:
\[ \mathbf{F}^\perp=\mathbf{F}-(\mathbf{F}\cdot \mathbf{\hat{N}})\mathbf{\hat{N}} \] 
\[ \mathbf{y'} = \mathbf{y} + \alpha \mathbf{F}^\perp \]
\[\mathbf{\hat{N}}'=\frac{\mathbf{y'-x}}{\vert \mathbf{y'-x} \vert}.\]
After a few steps the iteration is stopped before a locally optimal
lowest curvature mode is found. Initial velocities for the MD escape are then chosen
along the final escape direction $\hat{\mathbf{N}}$.

If the softening procedure is executed until it converges the
performance drops again. It is important not to overdo softening. Always
escaping into the {\em same} soft mode direction of a given minimum reduces the
possibilities of different escape directions and therefore weakens the
method. A good indicator was the mean kinetic energy during a run. For a few
softening iterations the value decreases whereas it starts to increase again
at a certain number of softening iterations. We set the iteration count
to the value where the mean kinetic energy was minimal.

Typically $40$ iterations are done with a step size $\alpha$
and a dimer length of $d=0.01$.

\paragraph{Enhanced Feedback.}
In original MH {\tt ekin} is increased by a factor $\beta_2$ if the current minimum
has already been visited before --- regardless of the number of previous visits. An
enhanced feedback method uses a value of $\beta_2$ depending on the previous visits
according to
\begin{equation}
\beta_2 = \beta_2^0 \times ( 1 + c \log{ N } )
\label{eqn:enhfb}
\end{equation}
where $\beta_2^0$ is the original value of $1.05$ and $N$ the number
of previous visits to this minimum. The parameter $c$ has been set to $0.1$ after
tests on bigger Lennard-Jones clusters and gold systems.
This feedback mechanism reacts slightly stronger if the minimum
is visited many times. 
If the system has only one energy funnel this enhanced feedback can even be slightly 
disadvantageous since it increases the kinetic energy too much and thus
weakens the BEP effect of MD. The increased feedback mechanism improves the
efficiency however considerably for large systems where the system can be trapped in 
huge structural funnels. If a cluster has for instance both low energy icosahedral and 
decahedral structures it takes a very long time for the MH algorithm without 
enhanced feedback to switch from one structure to the other. 

\paragraph{Parallel Minima Hopping.}
Parallelizing Minima Hopping is straightforward. On each processor an
own MH process is started, all sharing the same history list.  Only
energy values of visited minima have to be shared. Due to the feedback
mechanism and the common history list overlap of search areas is
penalized in this parallel setup and running on several processors can
thus yield an almost linear speedup in runtime. A further effect can
be exploited in parallel runs. A single run might easily get trapped
in a metastable funnel with a high escape time but the probability of
all processors getting trapped in this local minimum is exponentially
reduced with the number of processors running in parallel.  The total
expected runtime until success is therefore less influenced by the long
escape times of relatively stable local minima in parallel runs.  On the other hand
there is a minimal number of local minima that have to be visited before
the global optimum can be found. This minimal number of hops renders
the use of too many processors less efficient again thus leading to an
optimal number of processors depending on the structure of the PES.  The
idea of parallel sampling is known to have a positive effect~\cite{Shirts2001}.
%<+positive effects of parallel runs?? 2exp distr+>

\section{Computational Experiments on Clusters}

Atomic clusters are an ensemble of bound atoms, bigger than
a dimer but smaller than bulk matter. They show interesting
properties in the transition region of single atoms to bulk
matter. From a global optimization point of view they are complicated
multi-dimensional systems usually difficult to optimize since
they contain a lot of local minima.  Therefore they are of
interest to test the capabilities of an optimization algorithm.

%\begin{equation}
%E(\vert r_i - r_j \vert) = 4 \epsilon \left [ \left ( \frac{\sigma}{| r_i-r_j |} \right )^{12} - \left ( \frac{\sigma}{| r_i-r_j |} \right )^6 \right ],
%\label{eqn:lj}
%\end{equation}
A simple model of chemical interaction of two non-charged atoms is the
Lennard-Jones potential. The potential well depth and the equilibrium
distance are the only parameters, both are set to 1.

Such models represent rare-gas clusters reasonably well. Lennard-Jones
clusters are thoroughly studied model systems with well-known global
minima up to $1000$ atoms. Some of those clusters have a non-trivial
multi-funnel potential energy surface (PES) where the global minimum is
not easily found. The use of the Lennard-Jones two-body potential is
fast compared to more accurate potentials or even Density Functional
Theory (DFT) calculations. For these reasons they are well-suited
to test global optimization methods.  Our algorithms are applied to
Lennard-Jones systems consisting of $38$, $55$, $75$, $100$ and $150$
atoms. All clusters chosen show a different aspect, LJ$_{55}$ is a
very easy system, LJ$_{38}$ has a double funnel structure, LJ$_{75}$
is a very hard double-funnel system and LJ$_{100}$ and LJ$_{150}$
are examples of bigger clusters. Additionally, we perform single runs
on LJ$_{39}$, LJ$_{74}$ and LJ$_{98}$ to find the ground state motif
dependence of the algorithms. For these additional clusters we have
pairs of similar-sized clusters with very different ground state
geometries.

\begin{figure}
    \begin{center}
        \includegraphics[width=4cm]{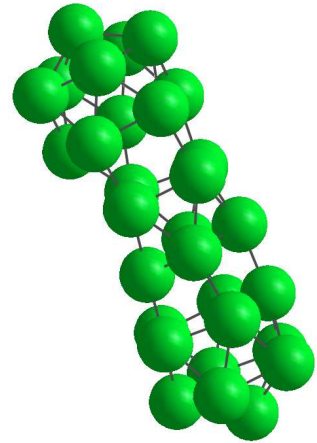}
    \end{center}
    \caption{Si$_{30}$ ground state with EDIP.}
    \label{fig:si30}
\end{figure}

Silicon clusters are of more practical interest as silicon is so widely
used in research and technology. To obtain realistic results usable
in those fields it would be necessary to calculate energies using at
least DFT methods.  But those methods are consuming a lot of CPU time
and one should highly optimize the algorithms applying DFT. Especially
in the field of global optimization it is of importance to use an
algorithm which is efficient to save computing time. To investigate
the behavior of the presented algorithm we use only a force field to
evaluate the energy of a configuration. Silicon systems are chosen as
model systems since they possess directed bonds and show frustrated
behavior.  They have non-trivial minimum structures which are in
general neither compact nor spherically symmetric. Silicon systems
containing $18$, $22$, $30$ and $60$ atoms are explored. The force
field used is Bazant's {\em Environment-Dependent Interatomic Potential
(EDIP)}~\cite{Bazant1996,Bazant1997,Justo1998,Goedecker2002}. This force
field has been chosen because it tends to elongated minimal structures
which are not spherical (FIG.~\ref{fig:si30}). There exist other force
fields preferring spherical ground states. But the performance with
spherical ground states is already investigated in Lennard-Jones and
gold systems.  Such highly elongated structures might not be very
realistic as more accurate DFT calculation suggest only slightly
elongated ground state configurations.

In gold systems different configurations can have very similar
energy and the global optimum is often only very slightly lower than
the next-best solution. This leads to a situation where the global
geometry of the gold cluster can change completely by adding only
a single atom. The energies of gold clusters are calculated using
an empirical many-body potential by Rosato, Guillopè and Legrand
(RGL)~\cite{Baletto2002}. Gold clusters with $28$, $76$ and $102$ atoms
are investigated. Additionally, Au$_{79}$ and Au$_{101}$ are added to
the test set in order to have different ground state geometries at
similar sizes.

\medskip
To compare different algorithms we measure two quantities, the total
number of calls to the function calculating energy and forces and the
total number of local geometry optimizations. The local optimization
is the most expensive step in both algorithms and the speed is mainly
determined by its number. To obtain meaningful numbers we perform
between $20$ and $100$ runs on each problem.
%iAs a remark to the number
%of function calls we point to the fact that in Minima Hopping runs,
%the calls to the energy function from within a MD escape step can be
%done using only low precision in DFT applications and therefore saving
%computer time.

The numerical convergence criterion for local optimizations is
$\Vert \mathbf{F} \Vert < 10^{-4}$ for the total euclidean force norm which leads to an energy
precision of almost $\sim 10^{-8}$.

\section{Results}

We gathered results for total performance on the investigated
systems.  Additionally, we also did comparative runs concerning the new
modifications mentioned above.  The numerical results of the different
performance runs can be found in TABLE~\ref{tbl:results} whereas special
test cases are addressed later.  The table shows the results obtained
using the best parameter set known to us. Those values are usually a
combination of both heredity methods and all mutations.  Minima Hopping
found the global minimum in all problems investigated whereas the
evolutionary algorithm could not find the global minima for LJ$_{75}$,
LJ$_{98}$, Si$_{30}$, Si$_{60}$, Au$_{79}$ and Au$_{102}$.

\subsection{General Performance}

The comparison shows a good applicability of Minima Hopping in all
problems.  The Evolutionary Algorithm is capable of finding most of the
ground states but fails in some cases. Where it is successful it can
even outperform MH.  Problematic for the EA are the non-icosahedral
ground states, such as LJ$_{75}$, LJ$_{98}$ and the elongated silicon
clusters.

%The comparison clearly states a stable superiority of Minima Hopping in
%this setup.  It is always best or at least almost as good as the EA. The
%evolutionary optimiser can maximally reach the performance of Minima
%Hopping or be slightly better in easy systems. For hard problems such
%as LJ$_{75}$ or the geometrically wild configurations of the bigger
%silicon clusters the evolutionary algorithm did not even converge to
%the globally best solution.  %Going to bigger clusters, EA seem to
%gain efficiency compared to %Minima Hopping. The series LJ$_{100}$ to
%LJ$_{160}$ show a strong %slowdown of MH.

The failure of the Evolutionary Algorithm to find the elongated
ground state structure of silicon systems shows a clear advantage of
Minima Hopping, it is not geometry dependent as the current Evolutionary
Algorithms are. Moving via MD it is applicable to any system for which
forces are available. In the past there have been approaches inspired by
genetic algorithms which used standard crossover and in the beginning
even binary strings as coordinate representation. Those ideas seem to
be a bit less geometry dependent than the nowadays applied heredity
methods.  But the original simple genetic methods have been shown
to be less effective than more elaborate geometrical operators in
Lennard-Jones systems~\cite{Pullan1997}.

When comparing two clusters of similar size but with different ground
state structures it is obvious that the evolutionary algorithm can
in general not reproduce non-icosahedral ground states with the
operators presented here.  The results of this tests can be found in
TABLE~\ref{tbl:structcomp}.  This table contains results of comparative
runs without enough statistics to compare performance.  The problem has
already been identified and solved by {\em niches} to prevent domination
of the whole population by only one geometry type~\cite{Hartke1999}.
Minima Hopping is able to find the non-icosahedral motifs without
further modification in the standard configuration though with a
decreased performance comparing to the icosahedral configuration of
similar size.

The results show that a further adaption of EA to some specific features of
clusters is necessary to enable the EA to work as efficiently for clusters
as they do for periodic solids.

A further difference can be noted when comparing function calls.
Minima Hopping tends to need in total more calls to the energy and
forces function than the evolutionary algorithm in the cases where
the EA succeeded, at least in smaller systems. This is clearly due to
the use of MD and softening in Minima Hopping. But we remark that it
is not necessary to perform MD escape with the full accuracy. A DFT
calculation can be done using a reduced basis set and therefore in
significantly less computer time. On the other hand Minima Hopping never
produces energetically awful candidates since it moves via MD. Since
heredity and mutation operators do not necessarily generate chemically
reasonable configurations the local geometry optimization requires more
force evaluations in the EA. Since in DFT applications the early stages of local
optimization do not need high precision they could even be done using
force fields thus reducing the computational demand.

\begin{table}
\centering
\begin{tabular}{l|l|r}
    \hline
    \hline
    {\bf System}      & {\bf Structure}       & {\bf Success} \\
    \hline
    LJ$_{39}$   & Icosahedron     & yes \\
    LJ$_{38}$   & fcc             & yes \\
    \hline
    LJ$_{74}$   & Icosahedron     &  yes \\
    LJ$_{75}$   & Marks decahedron & no  \\
    \hline
    LJ$_{100}$  & Icosahedron     & yes \\
    LJ$_{98}$   & Tetrahedron     & no  \\
    \hline
    Au$_{76}$   & Icosahedron     & yes \\
    Au$_{79}$   & fcc           & no  \\
    \hline
    Au$_{101}$  & Icosahedron   & yes \\
    Au$_{102}$  & fcc           & no   \\
    \hline
    \hline
    
\end{tabular}
\caption{Performance of the EA depends on the motif of the ground
state configuration. It is not possible to find non-icosahedral ground states in systems
bigger than 38 atoms. Minima Hopping was able to find all of the
listed configurations. A {\em no} means failure to find the ground state structure
within $100\,000$ local optimizations.}
\label{tbl:structcomp}
\end{table}

%%number are adapted to 10%MD accuracy and include softening, both via estimation
\begin{table*}
\begin{tabular}{l|r||r|r|r|r||r|r|r|r}
    \hline
    \hline
    \multicolumn{2}{l||}{{\bf Cluster}}    & \multicolumn{4}{c||}{{\bf Minima Hopping}} & \multicolumn{4}{|c}{{\bf Evolutionary Algorithm}} \\
        &   Energy  & \#GO$^1$  & calls (GO)$^2$ & calls (MD)$^3$ & runs$^4$  & \#GO  & calls & configuration$^5$ & runs\\
    \hline
    LJ$_{26}$   &$-108.31562$ & $96$        & $50\,610$        & $8\,300$          & $100$ & $\mathbf{56}$     & $\mathbf{34\,200}$         & $10-4-6$  & $100$ \\
    LJ$_{38}$   &$-173.92842$ & $1\,190$    & $688\,500$       & $85\,930$         & $100$ & $1\,265$          & $732\,900$                 & $25-0-20$ & $100$ \\
    LJ$_{55}$   &$-279.24847$ & $190$       & $74\,840$        & $14\,700$         & $100$ & $\mathbf{100}$    & $\mathbf{54\,900}$         & $10-3-3$  & $100$ \\
    LJ$_{75}$   &$-397.49233$ & $27\,375$   & $12.3\times10^6$ & $2.11\times10^6$  & $20$  & --                & --                         & --        & --    \\
    LJ$_{100}$  &$-557.03982$ & $5\,960$    & $1.87\times10^6$ & $417\,000$        & $42/50$  & $5\,908$       & $3.89\times10^6$           & $60-20-40$& $35/40$  \\
    LJ$_{150}$  &$-893.31026$ & $9\,490$    & $3.72\times10^6$ & $758\,000$        & $45/50$  & $\mathbf{7\,980}$ & $4.36\times10^6$        & $60-20-40$& $17/20$  \\
    \hline
    Si$_{11}$   &$-44.91223$ eV & $29$      & $11\,790$         & $4\,190$          & $60$  & $61$      & $31\,300$         & $10-3-4$  & $50$    \\
    Si$_{18}$   &$-74.88419$ eV & $110$     & $23\,100$         & $10\,700$         & $93$  & $195$     & $110\,100$        & $10-4-6$  & $40$    \\
    Si$_{22}$   &$-92.48090$ eV & $370$     & $187\,400$        & $52\,100$         & $21$  & $3\,300$  & $1.74\times10^6$  & $10-4-6$  & $1/5$   \\
    Si$_{30}$   &$-126.0952$ eV & $5\,050$  & $3.98\times10^6$  & $750\,000$        & $100$ & --        & --                & --        & --      \\
    Si$_{60}$   &$-253.05509$ eV& $23\,300$ & $14.1\times10^6$  & $2.80\times10^6$  & $44/49$  & --        & --                & --        & --      \\
    \hline
    Au$_{28}$   &$-99.95115$ eV & $54$      & $26\,210$         & $3\,510$          & $100$ & $87$      & $68\,910$            & $10-3-6$  & $50$ \\
    Au$_{76}$   &$-279.34791$ eV& $1\,124$  & $526\,500$        & $70\,800$         & $98$  & $2\,680$  & $1.60\times10^6$     & $25-5-13$ & $19/20$ \\
    \hline
    \hline
\end{tabular}
\caption{Best Results. Highlighted in bold face are the problems where
the EA performed better. All values are averaged over multiple runs,
indicated in the column {\em runs}. Standard deviations are as big as
the mean values. $^1$ Average number of local optimizations over all
runs. $^2$ Calls to energy function during local optimization including
softening. $^3$ Calls to energy function from MD escape steps. $^4$ If
two numbers are present some runs had to be stopped due to too long
runtime, the first number indicates the successful runs. $^5$ Configuration of EA: {\tt
populationsize} --- {\tt elitism} --- {\tt cutoff}.}
\label{tbl:results}
\end{table*}

\subsection{Heredity Methods}

While comparing both heredity methods used we observed that 
average offspring method performed better in systems with a compact
optimal structure. The
plane-cut method could only produce offspring as good as the
other method by choosing a very low {\tt cutoff}. When Deaven {\em et
al.}~\cite{Deaven1996} applied this method they produced every
single combination of two parents from a very small population
so having actually a very low cut-off level too. With the lower
cut-off level the {\em plane-cut} method performed well overall.

\begin{table}
\centering
\begin{tabular}{l||r|r||r|r}
    \hline
    \hline
    {\bf Cluster} & \multicolumn{2}{c||}{ {\bf Average Offspring}} & \multicolumn{2}{|c}{ {\bf Plane-Cut}} \\    
            & \#GO   & configuration & \#GO   & configuration \\
    \hline
    LJ$_{55}$   & $119$  &  $10-2-8$ & $\mathbf{100}$ & $10-3-3$ \\
    LJ$_{38}$    & $\mathbf{1265}$  & $25-0-20$  & $1595$ & $25-0-10$ \\
    Si$_{18}$    & $322$  & $10-3-6$  & $\mathbf{195}$ & $10-4-6$ \\
    Au$_{28}$    & $87$ & $10-3-6$  &  $88$ & $10-2-6$ \\
    \hline
    \hline
\end{tabular}
\caption{Heredity methods in direct comparison}
\label{tbl:compaopc}
\end{table}

\begin{figure*}
\centering
\subfloat[Average Offspring]{\includegraphics[width=0.4\textwidth]{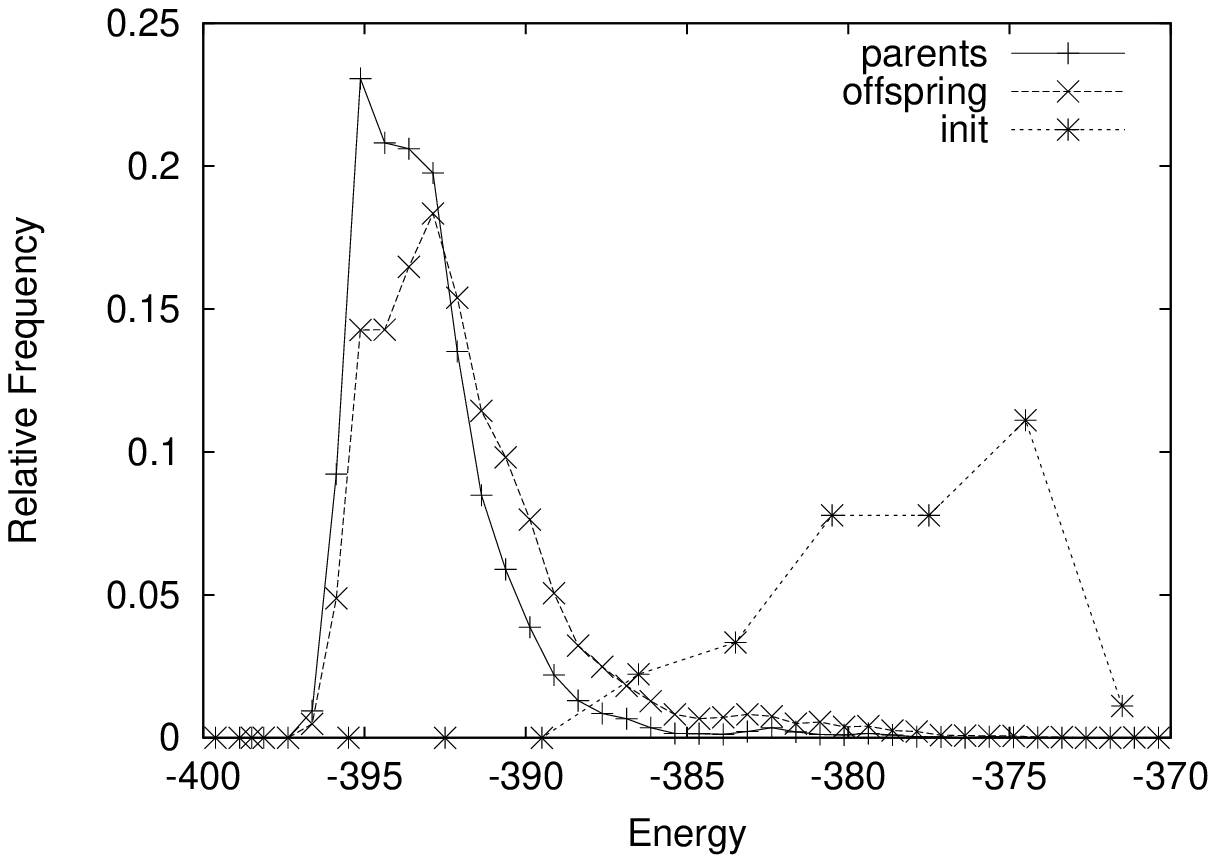}} 
\subfloat[Plane-cut]{\includegraphics[width=0.4\textwidth]{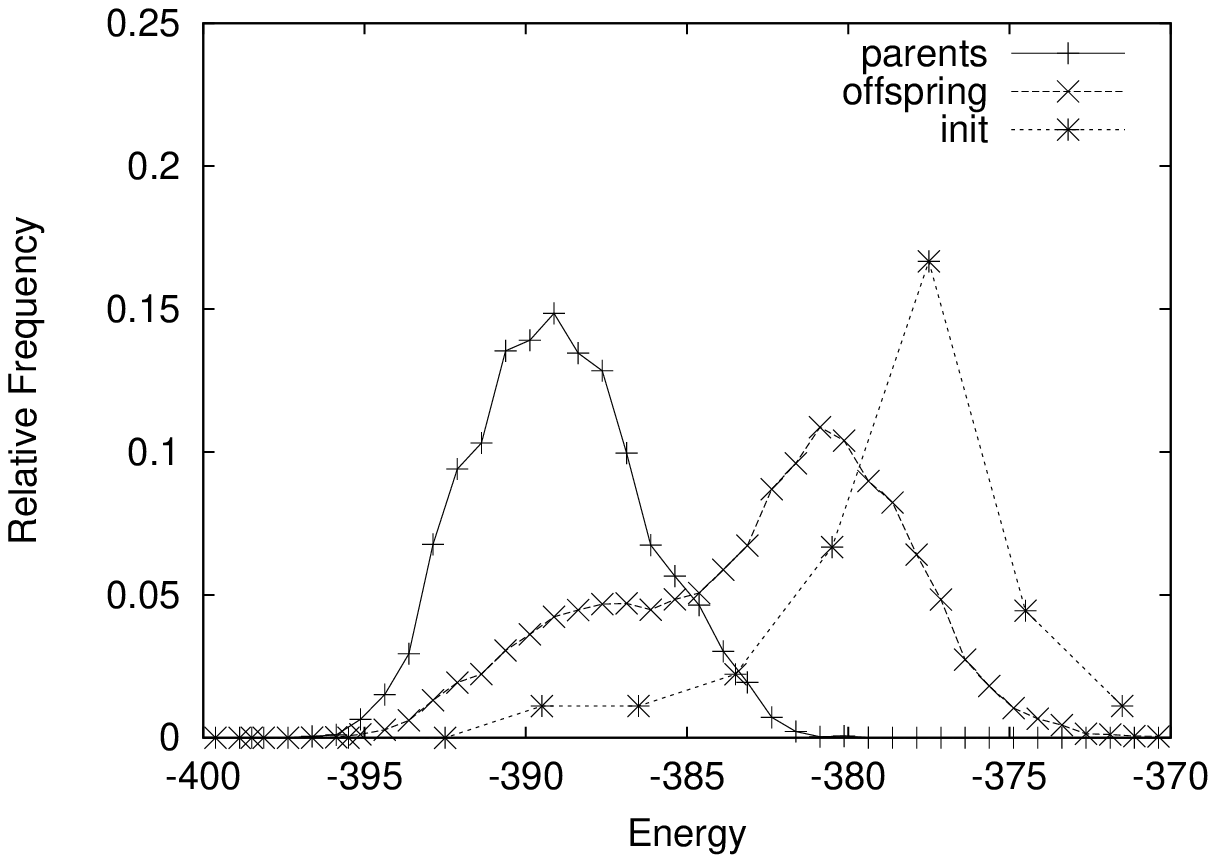}} \\
\subfloat[Combined 1:1]{\includegraphics[width=0.4\textwidth]{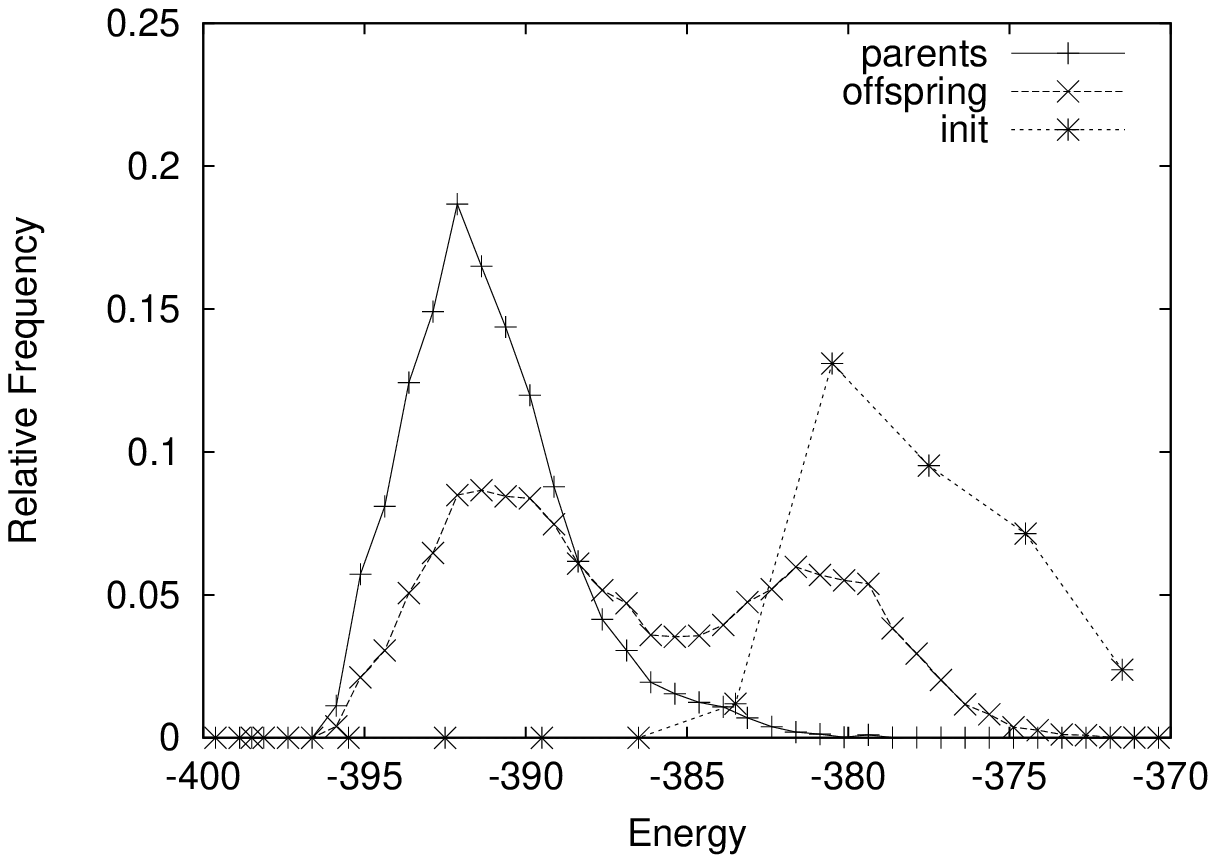}}
\caption{Candidate samples in LJ$_{75}$ after $10\,000$ minima. {\em
parents}: all energy values of candidates selected as parents, {\em
offspring}: all produced samples, {\em init}: initial population.
The global optimum is located at $-397.5$.
(a) {\em Average Offspring} method with a relative {\tt cutoff} of $1$,
(b) {\em Plane-Cut} method with a relative {\tt cutoff} $1/3$,
(c) Combined 1:1 with relative {\tt cutoff} $2/3$.}
\label{fig:samplehistaopc}
\end{figure*}

The samples obtained using {\em average offspring} method are
energetically closer to the fittest member of the population where the
{\em cutting plane} method samples broader with more diversity (FIG.
\ref{fig:samplehistaopc}.a/b).  It is a known fact that decreasing
diversity in the population can lead to premature convergence. On the
other hand, if samples in the complete energy range are allowed the
method resembles too much a random search and loses efficiency. To
observe a well evolving population it is necessary to have a balanced
distribution of selected individuals.  It seems therefore necessary to
lower the cut-off parameter in plane-cut runs to select more of the
fitter individuals as parents. A mix of both heredity methods delivered
the best results (FIG.~\ref{fig:samplehistaopc}.c).

The {\em plane-cut} method is better suited to a general application
of the algorithm, it can partially solve geometries less compact
than the {\em average offspring} method.

In big clusters it proved useful to combine all operators available. The
results, especially LJ$_{100}$, were best with a combination of all
presented methods. In general the mixture was at a $1:1$ ratio or even
more preferring {\em plane-cuts} in systems with known tendency towards
non-spherical ground state (e.g. Si$_{18}$ in TABLE \ref{tbl:compaopc}).
%<+Interpretation of the AO+>

We have also tested different plane-cut setups with a slightly modified
method where a minimal distance between the two cluster halves is
enforced.  This method performed poorly and was always weaker than all
different methods tested. Another modification where COM is not enforced
to lie in the plane was also considered and dropped since there were no
improvements.

The random rotation before recombination is necessary for {\em average
offspring} method but only of advantage for the {\em plane-cut} method
(see TABLE~\ref{tbl:randrot}).  If the system prefers non-spherical
configurations {\tt raterndrot} should be small.

\begin{table}
\centering
\begin{tabular}{l|r|r}
    \hline
    \hline
    {\bf Rotation}$^1$ & {\bf ~AO~} & {\bf ~PC~} \\
    \hline
    $0\%$  & $>1\,000$  &  $280$ \\
        $10\%$  & $343$  & $142$ \\
        $50\%$  & $174$  & $142$  \\
        $75\%$  & $123$  & $127$  \\
        $90\%$  &  $\mathbf{121}$ &  $\mathbf{115}$ \\
        $100\%$  & $130$  &  $147$ \\
        \hline
        \hline
    \end{tabular}
    \caption{Number of local optimizations for different random rotation
rates before heredity operator application for LJ$_{55}$
systems. For each setup $20$ runs were performed.
Random rotation is crucial for {\em average offspring} method (AO) 
but not for {\em plane-cut}
method (PC). $^1$ Frequency of random rotation.}
    \label{tbl:randrot}
\end{table}

\subsection{Parameter Tuning}

In systems with a double-funnel structure where the global maximum is
located in the narrower funnel it might turn out advantageous to disable
elitism, or (better) include in elitism only sufficiently different
structures. In our tests LJ$_{38}$ performed best when filling the
population with offspring only. But we should remark that average
offspring method has a rather preserving character often reproducing
candidates already known.

A drawback of the evolutionary algorithm is the need to tune many
parameters. A solution working with good performance on many different
systems without adjustment would definitely be of interest. Minima
Hopping performed better in this aspect, it never needed additional
tuning, all runs were done using the same set of standard parameters.
Using a standard set (TABLE~\ref{tbl:parameterlist}) for all problems
resulted in performance loss of the EA. The problems still converged
but with performances down to half of the tuned versions shown in
TABLE~\ref{tbl:results}.
A possible way to overcome this limitation is to fix the relative rates
of elitism and mutation etc. and only adjust {\tt populationsize} to
the specific problem.  Another possibility would be an automatically
self-adapting version which tunes the parameters during runtime. In this
case a stable and efficient scheme of parameter adaptation would be
needed which is clearly not a trivial task.

We note that for crystals with up 30-60 atoms in the unit cell, the
USPEX algorithm~\cite{Glass2006,Oganov2006} proved to perform very well
with essentially a universal set of parameters without any parameter
tuning. Evolutionary optimization of clusters, which are more complex
systems, is more sensitive to parameter values.

\subsection{Modification of Minima Hopping}

\begin{table}
    \centering
    \begin{tabular}{l|r|r|r|r}
        \hline
        \hline
         {\bf System}& {\bf No Softening}     & \multicolumn{3}{|c}{{\bf Softening}} \\
                     & $c=0$$^1$          &  $c=0$ & $c=0.1$ & $c=0.2$ \\
        \hline
        LJ$_{38}$   & $2\,062$  & $1\,217$       & $1\,190$          & $990$ \\
        LJ$_{55}$   & $320$     & $140$          & $190$             & $198$          \\
        LJ$_{100}$  & $9\,100$  & $7\,300$       & $4\,700$          & $5\,800$       \\
        LJ$_{150}$  & {\em n/a} & $14\,900$      & $9\,111$          & $11\,630$      \\
        \hline
        Au$_{28}$   & $167$     & $44$           & $44$              & $56$            \\
        Au$_{76}$   & {\em n/a} & $979$          & $890$             & $1024$         \\
        \hline
        \hline
    \end{tabular}
    \caption{Geometry optimizations with and without {\em softening} in
    Minima Hopping with different feedback parameters. {\em n/a}: not
    tested.  $^1$ Parameter $c$ is defined in eq. (\ref{eqn:enhfb})}
    \label{tbl:soften}
\end{table}

Minima Hopping has been considerably improved using {\em softening}, in
all studied cases.

The use of enhanced feedback method is advantageous in large or
multi-funneled systems but can even have a negative effect in easy
systems as LJ$_{55}$ (TABLE~\ref{tbl:soften}).  The parameter $c$ in
(\ref{eqn:enhfb}) should not be chosen too large.  We used softening and
enhanced feedback with $c=0.1$ in the comparison runs.

\section{Conclusion}

We have tested an Evolutionary Algorithm capable of finding ground
state structures of atomic clusters.  In spite of the success of EA for
periodic systems and on surfaces the current Evolutionary Algorithm is overall
less efficient than Minima Hopping in the current implementation. It
is not yet able to find global minima with geometrically difficult
structures such as elongated silicon clusters and non-icosahedral ground
states without the concept of niches. In contrast, Minima Hopping was
able to find all ground states. Where the EA succeeds its performance is
comparable to or even better than that of MH.

Further improvements of the EA could make it superior to MH for cluster
optimization if the specifics of cluster systems are taken into account,
as it was already done for periodic systems in the USPEX algorithm.  The
MH algorithm, on the other hand, shows a high stability and little need
for further adaptation, at least for homoatomic systems.

Minima Hopping was improved by doing escape steps in directions with
relatively low curvature of the PES and by using an enhanced feedback
method.

%\bibliographystyle{apsrev}
%\bibliography{evolutionaryoptimisation}

\end{document}